\documentclass[prb,showpacs,twocolumn,superscriptaddress]{revtex4}
\usepackage{hyperref,graphicx,dcolumn}
\usepackage{amsfonts,amsmath,amssymb,bm,bbm}
\usepackage{color}
\usepackage{pdfsync}
\usepackage{blkarray}
\usepackage{multirow}
\usepackage{braket}
\usepackage{lipsum}

\newcommand{\myFig}[6]{ %
\begin{figure}[htb] 
\begin{center} 
\includegraphics[width=#1\columnwidth,height=#2\columnwidth,clip=true,keepaspectratio=#3]{#4}
\caption{#5} \vspace{-0.5cm} \label{#6} 
\end{center} \end{figure}}

\newcommand{\myket}[1]{\left\vert #1 \right\rangle}
\newcommand{\Avr}[1]{\left\langle #1 \right\rangle}

\newcommand{\Tr}[1]{ \mathrm{Tr} \left[ #1 \right]}
\newcommand{\SqB}[1]{ \left[ #1 \right]}
\newcommand{\RnB}[1]{ \left( #1 \right)}

\begin{document}
\title{The role of spin-orbit interaction in the ultrafast demagnetization of small iron clusters}
\author{Maria Stamenova}\email[Contact email address: ]{stamenom@tcd.ie}
\affiliation{School of Physics, AMBER and CRANN Institute, Trinity College, Dublin 2, Ireland} 
\author{Jacopo Simoni} 
\affiliation{School of Physics, AMBER and CRANN Institute, Trinity College, Dublin 2, Ireland} 
\author{Stefano Sanvito} 
\affiliation{School of Physics, AMBER and CRANN Institute, Trinity College, Dublin 2, Ireland} 

\begin{abstract}
The ultra-fast demagnetization of small iron clusters initiated by an intense optical excitation is studied with time-dependent spin density functional theory (TDSDFT). In particular we investigate the effect of the spin-orbit interaction on the onset of the demagnetization process. It is found that the initial rate of coherent spin loss is proportional to the square of the atomic spin-orbit coupling constant, $\lambda$. A simplified quantum spin model comprising spin-orbit interaction and a local time-dependent magnetic field is found to be the minimal model able to reproduce our {\it ab initio} results. The model predicts the $\lambda^2$ dependence of the onset rate of demagnetization when it is solved either numerically or analytically in the linear response limit. Our findings are supported by additional TDSDFT simulations of clusters made of Co and Ni.
\end{abstract}


\maketitle

Achieving control over the magnetization dynamics at the femtosecond timescale is a desirable 
asset for new magnetic data storage technologies. The ultrafast optical demagnetization (UOD) phenomenon, 
discovered by Beaurepaire {\it et al} \cite{First96}, in which an intense femtosecond laser pulse 
induces an abrupt and dramatic loss of magnetization in a metallic film, initiated what is now 
the highly active field of femto-magnetism. 
Typical UOD experiments are based on the pump-probe method, where a 
femtosecond laser pulse in the optical range (pump) is shed onto the magnetic sample and 
then a delayed short electromagnetic pulse (probe) is used to detect the magnetic response through 
possible linear or non-linear magneto-optical effects \cite{Rasing10}. By varying the time delay 
between the pump and the probe the magnetization dynamics can be reconstructed in the time domain 
over a typical range spanning from a few femtoseconds to a few picoseconds. The rapid demagnetization process 
that develops over this time can be characterized by two distinct 
stages: (i) a coherent stage in the first few tens of fs when the light interacts with the electrons and (ii) a relaxation stage when 
hot electrons and spins interact with each other and with the lattice so to thermalize. Although the role 
of the particular microscopic spin-flip mechanisms is often unclear and dependent on the details of the
magnet investigated, the thermalization process is in general tractable through empirical three-temperature 
models \cite{First96,Koopmans10}, which establish rate equations between the spin, electron and phonon systems. 
In contrast, theory for the coherent stage is rather unsettled and spans a range of different views (not necessarily mutually exclusive), from relativistic accounts of the direct photon-spin coupling \cite{Hinschberger12} to 
semi-classical transport models \cite{Battiato10}.  

Experimental works, focussed on the coherent regime, have described a strong dependence of the rate of UOD on the material and, particularly, its spin-orbit coupling (SOC) properties. For instance, it has been  
reported \cite{Bigot09} that materials exhibiting stronger SOC demagnetize significantly faster than 
lighter ones. The SOC has been identified as a key component enabling UOD also in earlier theoretical works based on model Hamiltonian \cite{Zhang00}. Very recently time-dependent spin-density functional theory (TDSDFT) calculations \cite{Krieger15} have provided another confirmation of its essential role for the ultrafast laser-induced loss of spin in bulk transition metals. In this letter we seek to gain further understanding of the microscopic mechanisms responsible for the very initiation of the UOD. We employ the only practically-applicable first-principles theoretical framework, the TD(S)DFT \cite{TDDFTGross2,SpinDynQian}, that allows to simulate the UOD process for real atomic clusters directly in the time domain and for experimentally-relevant times. We demonstrate that the very onset of the demagnetization is triggered by the electronic charge response to the electric field of the pulse. The charge and spin currents generated give rise to a magnetic field which in combination with the SOC facilitates spin flips and global spin decay. We also establish that the initial, coherent demagnetization rate is proportional to the square of the ionic SOC strength for a range of small transition metal clusters.

\myFig{0.8}{0.8}{true}{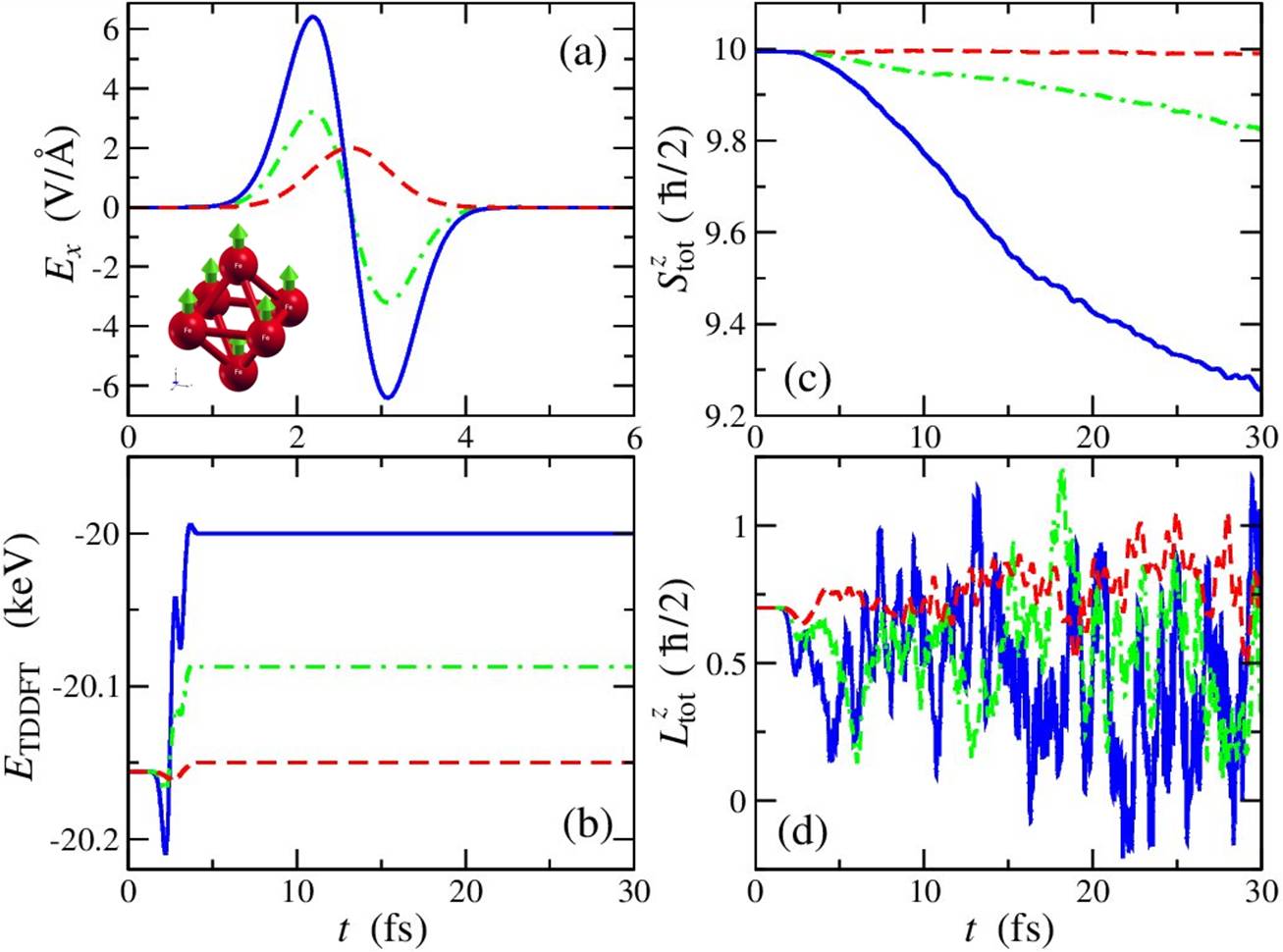}{(Color online) (a) Typical electric field pulses used to excite the Fe$_6$ cluster 
(cartooned as inset) and time evolution of the total TDSDFT energy (b), total spin (c) and total KS angular momentum (d) 
of the cluster when subjected to each of the pulses in panel (a) with corresponding color code on each of the other
panels. }{fig01}

In particular, we focus on Fe$_6$, the geometry and ground state (GS) spin $2S=20\,\hbar$ of which have been previously predicted 
by the LSDA\cite{Gutsev}. It is well-known that open $d$-shell systems are problematic to local
approximations of the exchange and correlation (XC) functional. Here, however, our intention is to study the 
generality of the spin dynamics so that possible quantitative features are not important at this time. 
We consider the adiabatic temporal extension \cite{ALDA} of the local spin-density approximation (ALSDA), 
parameterized by Perdew and Wang \cite{PW} and implemented in the Octopus code \cite{Octop1}. 

In all our calculations spin dynamics is initiated by a single intense electric field pulse (we neglect the 
magnetic field component). We solve the time-dependent Kohn-Sham (KS) equations	
\begin{equation}
i\hbar\frac{\partial}{\partial t}\psi_{j}(\vec{r},t) = H_\mathrm{KS}(\vec{r},t)\psi_{j}(\vec{r},t)\:,
\end{equation}
where $H_\mathrm{KS}(\vec{r},t)$ is $2\times 2$ matrix in spin space to account for non-collinearity, 
$\psi_{j}$  are two-dimensional spinors and we consider the time-dependent KS Hamiltonian 
with electric field, $\vec{E}(t)$, introduced in the length gauge
\begin{eqnarray}
H_\mathrm{KS}(\vec{r},t) & = & -\frac{\hbar^{2}\nabla^{2}}{2m} + v_\mathrm{s}(\vec{r},t) - \frac{2\mu_\mathrm{B}}{\hbar}\hat{\vec{S}}\cdot\vec{B}_\mathrm{xc}(\vec{r},t) \,\,\,\,\,\, \\
v_\mathrm{s}(\vec{r},t) & = & \sum_{\vec{R}_{I}} V_\mathrm{PP}(|\vec{r}-\vec{R}_{I}|) + \int d^{3}r' \frac{n(\vec{r'})}{|\vec{r}-\vec{r'}|} + \nonumber \\ & & v_\mathrm{xc}(\vec{r},t) + \vec{r}\cdot \vec{E}(t). 
\end{eqnarray}
We substitute the ionic potentials, centered at each site $\vec{R}_{I}$, with soft norm-conserving fully relativistic pseudopotentials that 
reproduce correctly the semi-core and valence wavefunctions beyond a certain core 
radius\cite{PachecoSemicore, AbInitSOC,APE} \footnote{Our fully-relativistic norm-conserving pseudopotentials 
have been generated with the multi-reference pseudopotential method as implemented in the APE 
code \cite{APE}. This includes, in addition to valence states, also the semi-core ones ($3s$ and 
$3p$), since semi-core states play an important role in describing the chemical bond and the 
magnetic properties of transition metal clusters \cite{Wang92}.}
\begin{eqnarray} \label{eq:Vpp}
V_\mathrm{PP}(r) & = & \sum_{l}\sum_{m=-l}^{l}\bigg( \bar{V}_{l}^\mathrm{ion}(r) + \frac{1}{4} V_{l}^\mathrm{SO}(r) + \nonumber \\ 
 & &  V_{l}^\mathrm{SO}(r) \hat{\vec{L}} \cdot \hat{\vec{S}}\bigg)\ket{l, m}\bra{l, m}\:.
\end{eqnarray}
Here, $\hat{\vec{S}}$ is the spin operator, $\hat{\vec{L}}$ is the angular momentum operator 
associated to the atomic center, while the scalar 
part of the pseudopotential $\bar{V}_{l}^\mathrm{ion}(r)$ includes the effect of the mass shift and the 
Darwin term. $V_{l}^\mathrm{SO}(r)$ defines the range of the SOC term 
\footnote{For values of the radius greater than the range of the SOC component the ionic 
pseudopotential reduces to a purely local object.}.  Within the adiabatic LSDA, at every 
time $t$ the XC potentials are calculated through the ground state LSDA exchange-correlation 
energy functional corresponding to the instantaneous electron charge density 
$n(\vec{r},t)=\sum_{j\in \mathrm{occ.}} \psi_j^*(\vec{r},t)  \psi_j(\vec{r},t)$ and spin density 
$\vec{s}(\vec{r},t)=\frac{\hbar}{2}\sum_{j\in \mathrm{occ.}} \psi_j^*(\vec{r},t) \vec{S}  \psi_j(\vec{r},t)$ , i.e.
\begin{equation}
v_\mathrm{xc}(\vec{r},t) \!=\!\left. \frac{\delta E_{xc}}{\delta n}\right|_{\!\begin{array}[b]{c} n(\vec{r},t)\\ \vec{s}(\vec{r},t) \end{array}}, \,\, \vec{B}_\mathrm{xc}(\vec{r},t) \!=\!\frac{\hbar}{2\mu_\mathrm{B}}\left. \frac{\delta E_{xc}}{\delta \vec{s}}\right|_{\!\begin{array}[b]{c} n(\vec{r},t)\\ \vec{s}(\vec{r},t) \end{array}} .
\end{equation}
For the XC magnetic field the zero-torque theorem holds \cite{SpinDynTDDFT}, i.e.
$\int d^{3}r \vec{B}_\mathrm{xc}(\vec{r}) \times \vec{s}(\vec{r}) = 0$,
where the integral is taken over the entire space. In other words, $\vec{B}_\mathrm{xc}$ cannot produce 
a global spin torque over the system. The Heisenberg equation of motion for the spin operator leads to a spin 
continuity equation in the form
\begin{equation} \label{eq:spincont}
\frac{d\vec{s}(\vec{r},t)}{dt}  =  -\nabla\cdot \mathrm{J}_\mathrm{KS}(\vec{r},t) + 2\frac{\mu_\mathrm{B}}{\hbar}\vec{s}(\vec{r},t)\times\vec{B}_\mathrm{xc}(\vec{r},t) + \vec{\Gamma}(\vec{r},t) 
\end{equation}
where
\begin{eqnarray}
\vec{\Gamma}_{n}(\vec{r},t) & = & \sum_{\vec{R}_{I}}\sum_{\alpha,\beta}\sum_{m,m',l} \epsilon_{njk}V_{l}^\mathrm{SO}(|\vec{r}-\vec{R}_{I}|)\cdot \nonumber \\
& & \cdot \braket{\Psi^\mathrm{KS}_{\alpha}|l, m', \vec{R}_{I}} \hat{S}_{\alpha,\beta}^{k}  \label{eq:3}  \\
& & \bra{l, m', \vec{R}_{I}}\hat{L}^{j}_{I}\ket{l, m, \vec{R}_{I}} \braket{l, m, \vec{R}_{I}|\Psi^\mathrm{KS}_{\beta}}. \nonumber
\end{eqnarray}

The first term in Eq. (\ref{eq:spincont}) can be written as a two components tensor, 
$ \mathrm{J}^{ij}_\mathrm{KS}(\vec{r},t)=  \Tr{\hat{S}^i \sum_k (\psi_k^* \partial_{j} \psi_k - \psi_k \partial_{j} \psi^*_k)}$ 
is the KS spin-current operator. The second term is the torque exerted locally by $\vec{B}_\mathrm{xc}(\vec{r},t)$, 
which vanishes within the adiabatic LSDA ($\vec{B}^{ALDA}_\mathrm{xc}(\vec{r})\parallel \vec{s}(\vec{r})$). 
The first term, however, can be re-written in a form $ \frac{2\mu_\mathrm{B}}{\hbar}\vec{s} \times \vec{B}_\mathrm{kin}$, 
where the {\it kinetic} 
magnetic field\cite{Antropov03} is defined as $\vec{B}_\mathrm{kin} = \frac{1}{2en} \sum_k \nabla_k (n \nabla_k \vec{s})$. 
An effective local magnetic field  can be defined as 
$\vec{B}_\mathrm{eff} (\vec{r},t)=\vec{B}_\mathrm{kin} (\vec{r},t) + \vec{B}_\mathrm{xc} (\vec{r},t)$. 
This field is not necessary locally parallel to $\vec{s} (\vec{r},t)$ and produces a local torque. 

The last term in Eq. (\ref{eq:spincont}) $\vec{\Gamma}(\vec{r},t)$ is the only source of global spin relaxation 
in the temporal evolution. Note that in this description the global spin change is only determined by the orbital 
dependent scattering properties of the non-local component of the atomic pseudopotential as a result of the 
SOC and it is, therefore, not dependent directly on the Kohn-Sham orbital momentum defined 
as $\vec{L}_\mathrm{KS}=\vec{r}\times\vec{J}_\mathrm{KS}$.  Later in this letter we will revisit 
Eq.~(\ref{eq:spincont}) and use it as a base for a simplified quantum model for the spin operator. 

\myFig{1}{1}{true}{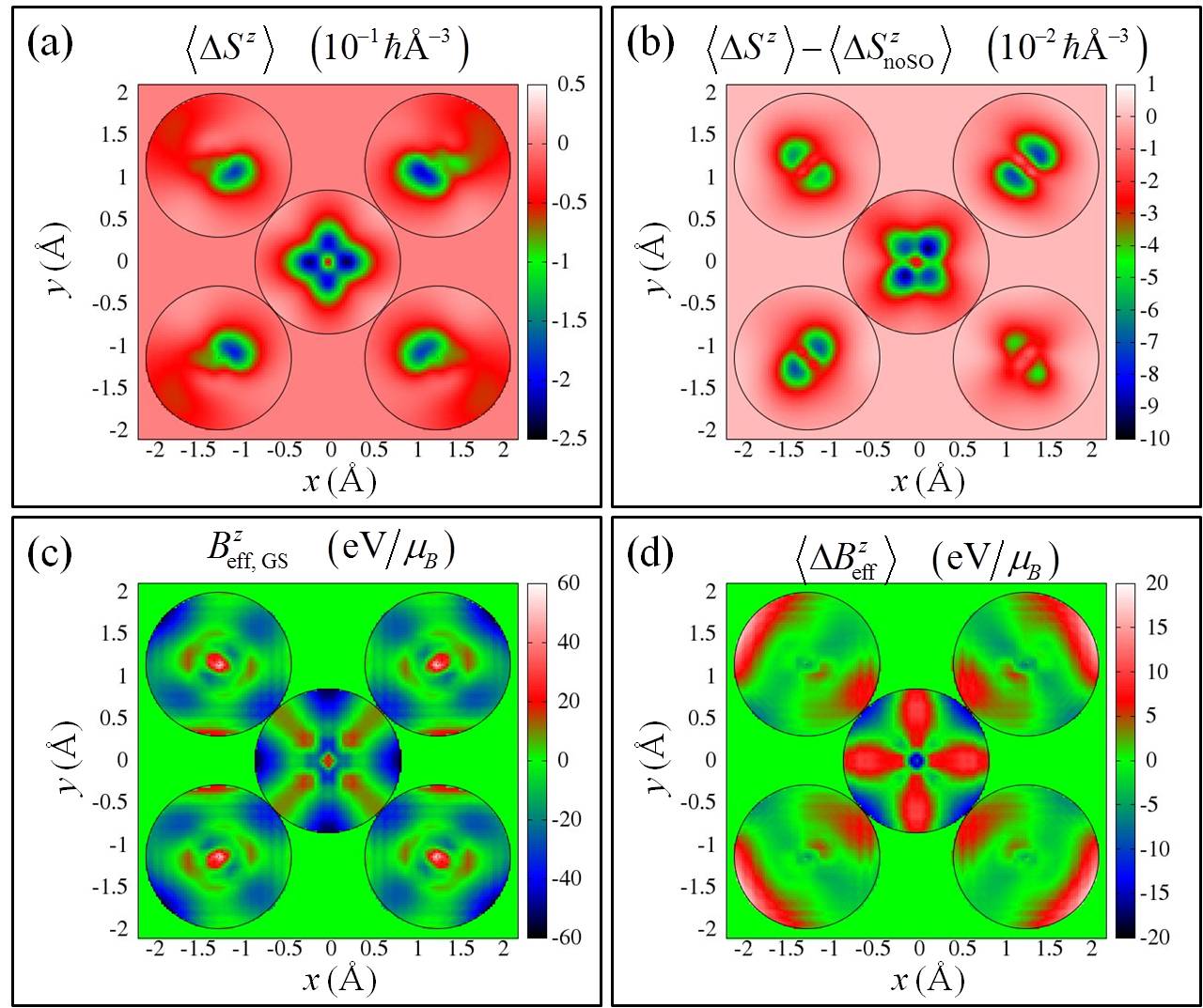}{(Color online) Contour plots of the time-averaged and $z$-averaged observables 
evaluated only in spheres of radius 0.85~\AA\ around each atom: (a) the averaged temporal variation of the 
spin density $S^z (\vec{r})$ with respect to the ground state; (b) difference of the latter and its counterpart in 
the case of no SOC; (c) the effective magnetic field $B^z_\mathrm{eff} (\vec{r})$ in the ground state 
($t=0$) and (d) the variation of $B^z_\mathrm{eff}(t)$ with respect to the ground state. 
See text for details.}{fig02}

Figure~\ref{fig01} shows representative results from the full time-dependent simulations of the dynamics
of the Fe$_6$ cluster. For a range of pulse shapes and amplitudes the magnetic response is a decay in the 
global spin expectation value, $S^z (t) = \int d^3 r s^z(\vec{r},t)$ . The pulse excites the cluster and the rate of 
the spin decay triggered by the excitation is correlated with the total variation of the TDSDFT energy 
before and after the pulse (the larger the energy deposited in the cluster, the larger the induced spin-decay 
rate). Note, that after the pulse the total energy is conserved, Fig.~\ref{fig01}(b). 

The spatial distribution of the calculated demagnetization is visualized in Fig.~\ref{fig02}, where we plot 
the time and space averaged (along the direction of the symmetry axis of the cluster, $z$) planar distributions of the 
temporal variations (with respect to the GS) of the spin-density and the effective magnetic field 
\footnote{The definitions of the quantities plotted in Fig.~\ref{fig02} are:   
$\overline{\Delta S^z(x,y)} = \frac{1}{T} \sum_i \int_T dt \int_{z_{i,1}}^{z_{i,2}} dz $ $ \SqB{s^z \RnB{x,y,z,t} - s^z \RnB{x,y,z,0}} / \RnB{z_{i,2}-z_{i,1}}$ 
and  $\overline{\Delta B_\mathrm{eff}^z(x,y)}\, =\, \frac{1}{T} \sum_i \int_T dt \int_{z_{i,1}}^{z_{i,2}} dz $ $  \SqB{B_\mathrm{eff}^z \RnB{x,y,z,t} -  B_\mathrm{eff}^z \RnB{x,y,z,0}}\,/\,\RnB{z_{i,2}-z_{i,1}}$,  where $T$ is the total simulation time and $z_{i,1}, z_{i,2}$ are functions of $(x,y)$ and belong to one of the 
non-overlapping atom-centered spheres $\Xi_i$.}.
It is notable that the negative variation of $s^z(\vec{r})$ is predominantly localized around the atomic 
centers [panel (a)]. Furthermore, the averaged spin-variation distribution difference between analogous 
simulations with and without SOC [panel (b)], which approximately represents the global spin loss 
in the presence of the SOC, is also localized\footnote{The apparent symmetry breaking is due 
to the finiteness of the time domain simulation.}. In particular it is more pronounced along the direction 
of the bond with the apex atoms, the shortest bond length in the system along the direction of the laser 
electric field. The effective magnetic field is very inhomogeneous in the GS. Its temporal variation, 
however, shows a spatial correlation with the variation of the spin density. The regions of decrease 
of spin density exhibit an increase in $B^z_\mathrm{eff}(\vec{r})$ (note that in the same regions 
$B^z_\mathrm{eff}(\vec{r})$ is mostly negative in the GS). Hence, regions of spin decay are associated 
with decrease in the absolute value of $B^z_\mathrm{eff}(\vec{r})$.

\myFig{0.9}{0.9}{true}{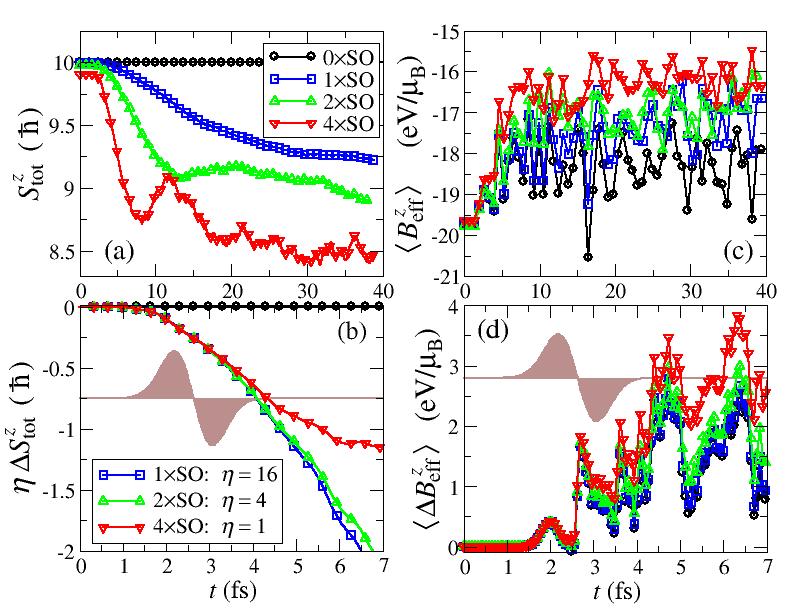}{(Color online)  (a) Evolution of global spin expectation value for different 
factors $\alpha$ in front of the SO term in Eq. (\ref{eq:Vpp}). (b) Same as in panel (a) but having the 
GS spin subtracted and $\Delta S^z_\mathrm{tot}$ values multiplied by $(1/\alpha^2)$. Panels (c) 
and (d) show the corresponding trajectories of the averaged over the non-overlapping atom-centered 
spheres $B^z_\mathrm{eff}(t)$ [see Eq (\ref{eq:Bxcloc})].}{fig03}

As suggested by Eq. (\ref{eq:spincont}), the SOC is expected to have a major role in the
spin-decay process. In order to extract such effect we have introduced an artificial scaling factor, $\alpha$, 
in front of the SOC term of Eq.~(\ref{eq:Vpp}). Depicted in Fig.~\ref{fig03} is the effect on the global 
spin-variation trajectory of the variation of $\alpha$ from 0 to 4. The rate of spin loss, both pulse-coherent 
and post-pulse, is strongly affected by the SOC strength with the limit of $\alpha=0$ (no SOC) resulting in global 
spin conservation. In panel (b) we have plotted the same spin trajectories after removing their GS offset and 
scalling them by a factor of $1/\alpha^2$. The overlap of the curves demonstrates that in the initial coherent stage the 
spin-decay rate scales as the square of the SOC strength. 

We now compare the global spin trajectories to those of the $B_\mathrm{eff}^z$ in the vicinity of the 
atomic centers. We define a measure of the local variation of the latter as
\begin{equation} \label{eq:Bxcloc}
\Avr{B^z_\mathrm{eff}(t)} = \sum_i \frac{1}{V_{\Xi_i}} \int_{\Xi_i} B_\mathrm{eff}^z (\vec{r},t) d^3 r \,, 
\end{equation}
where $\Xi_i$ are non-overlapping atom-centered spheres of radius 0.85 \AA\, for all the quantities plotted 
in~\ref{fig03}(c,d). Although such defined $\Avr{B^z_\mathrm{eff}(t)}$ appears noisy due to spatial grid effects, it does show 
a coherent response to the external field pulse and during this stage it is practically independent of the SOC strength. 
After the pulse dies out the decrease 
in the absolute value of $\Avr{B^z_\mathrm{eff}(t)}$ correlates to the global spin decay in panel (a), 
and that is especially notable for higher $\alpha$. 
This is related to the fact that practically all the spin loss takes place in the same atomic vicinity regions 
where $\Avr{B^z_\mathrm{eff}(t)}$ is defined. 

\myFig{0.9}{0.9}{true}{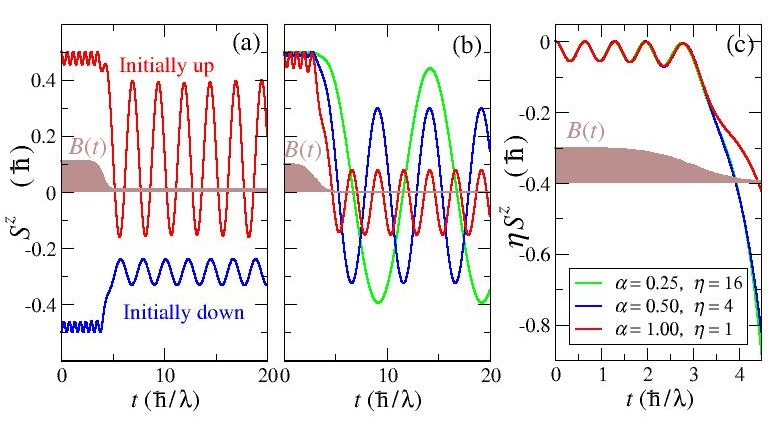}{(Color online) Trajectories for $S^z$ corresponding to the model Hamiltonian of
Eq.~(\ref{eq:modH}) in the case of (a) having an initial state with spin up or spin down or (b) different values of the SOC (factor $\alpha$) for an initial spin up. (c) Similarly to Fig.~\ref{fig02}(b), the latter 
trajectories re-scaled by $1/\alpha$. The shaded area is a reference for the temporal profile of $\vec{B}=[0,0,B(t)]$. 
The initial orbital momentum state is a linear combination of $l^z=0,1,2$ states.}{fig04}

\myFig{0.9}{0.9}{true}{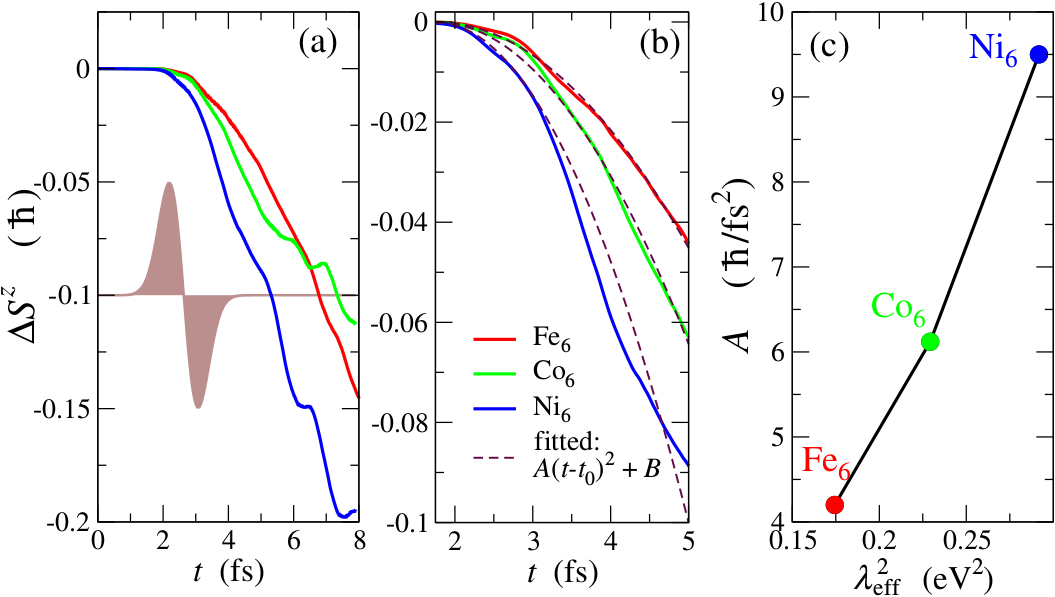}{(Color online) (a) TDSDFT trajectories of $S^z_{tot}$ with respect of the GS for 
three different clusters: Fe$_6$, Ni$_6$ and Co$_6$, all sharing the same Fe$_6$ geometry \cite{Gutsev}. 
(b) Closer view into the coherent part of the trajectories where a parabolic decay [$y=A(t-t_0)^2 +B$] is fitted 
for each trajectory (dashed curves). (c) The coefficient in the panel (b) fit $A$ versus an effective atomic 
SOC [Eq. (\ref{eq:lambdaeff})] for each material.}{fig05}

The insights drawn from Fig.~\ref{fig03} for Fe$_6$ suggest that the SOC is key in the demagnetization 
process, which in turn takes place in the vicinity of the atomic sites, i.e. where the SOC is the strongest. 
Furthermore, we have observed that in the same regions the $B_\mathrm{eff}^z$ also decays rapidly in time, 
coherently with the laser field.
As a minimal model for understanding the demagnetization process we propose the following spatially 
homogeneous and time-dependent spin Hamiltonian
\begin{equation}\label{eq:modH}
\hat{H}(t) = \lambda \hat{L}\cdot\hat{S} + \vec{B}(t)\cdot\hat{S}\:,
\end{equation}
where $\lambda$ defines the SOC strength and $\vec{B}(t)$ is a time-dependent magnetic field. 
This model effectively mimics the local effective spin dynamics at a given point in space arising from the TDSDFT 
calculation as described by Eq.~(\ref{eq:spincont}). The basis set used to expand the wavefunction, solution of 
the corresponding time-dependent Schr\"odinger equation, is given by the eigenstates of $\hat{L}^{z}$ and 
$\hat{S}^{z}$, $\ket{l^z, s^z}$. For instance, considering $l=1$ for the orbital momentum quantum number we 
can write
\begin{equation}
\ket{\Psi(t)} = \sum_{l^z=-1}^{1}\sum_{s^z=-1/2}^{1/2} c_{l^z,s^z} \ket{l^z, s^z}\:,
\end{equation}
and solve numerically the 6-dimensional Schr\"odinger equation to obtain the evolution of the spin observables. 
The calculated dynamics show that in the absence of SOC there is no spin dynamics regardless
of whether or not the initial state is collinear to $\vec{B}(t)$. This is because the spin and orbital angular momenta 
are decoupled. 

In contrast when $\lambda\neq 0$, an initial state with $l^z \ne \pm l$  and a step-like variation of $\vec{B}(t)$ 
(similarly to $\Avr{B^z_\mathrm{eff}(t)}$ in Fig. \ref{fig03}) produce a sharp change in the expectation value 
of $\hat{S}^z$ (see Fig. \ref{fig04}). In particular, for an initial spin-up state we find a decrease of the $S^z$ 
expectation value, while an initially down spin-state shows an increase in $S^z$. In other words, any change 
of the local magnetic field, combined with the SOC, leads to a decrease in the modulus of the 
expectation value of the spin of even initially collinear with the field (assumed along the quantization axis) spin-states with $l^z \ne \pm l$ 
(like the states in the open $d$-shell of the Fe atoms). 
In addition, the model reproduces the $\lambda^2$ dependence of the coherent demagnetization rate 
observed in the TDSDFT calculations [Fig. \ref{fig03}(b)]. In Fig. \ref{fig04}(b) and (c) we show that 
if the SOC strength is rescaled by a factor $\alpha$ the demagnetization curves get steeper
by a factor $\alpha^2$. 
%
%
This property can also be demonstrated analytically by looking at the first order term in the perturbative 
expansion of the solution $\myket{\Psi(t)}$. For instance, in the case of 
$\myket{\Psi(0)} = c_{1}(0) \myket{1,-1/2} + c_{2}(0)\myket{0, 1/2}$, the variation of 
expectation value of $\hat{S}^z$ with respect to the ground state reads
\begin{equation}\label{eq:deltaSz}
\Braket{\hat{S}^{z}(t)} - \Braket{\hat{S}^{z}}_{0} = \frac{\lambda^{2}t^{2}}{2}\big( -\frac{\sqrt{2}}{2}c_{1}c_{2} - c_{1}^{2} + c_{2}^{2}\big)\:,
\end{equation}
i.e. it scales as $\lambda^2$.

As a final proof for the $\lambda^2$ dependence of the demagnetization speed, we look at the laser-induced 
response of clusters analogous to Fe$_6$ but composed of Co and Ni (we keep the same geometry of Fe$_6$). 
We quantify their ionic SOC strength through the following definition
\begin{equation}\label{eq:lambdaeff}
\lambda_\mathrm{eff} = \sum_{l \in \mathrm{occ.}} \frac{n_l}{n_\mathrm{tot}} \int V^\mathrm{SO}_l (\vec{r}) R_l^2(\vec{r}) d^3 r \:,
\end{equation}
where $n_l$ are the KS state occupations with $l$ spanning the valence states (in this case $3s, 3p, 3d$ and $4s$),
$n_\mathrm{tot}= \sum_{l \in \mathrm{occ.}}n_l$, $R_l(\vec{r})$ are the radial pseudo-atomic wavefunctions and 
$V^\mathrm{SO}_l(\vec{r}) = \frac{2l}{2l+1}\SqB{V_\mathrm{PP}^{l+1/2}(\vec{r})-V_\mathrm{PP}^{l-1/2}(\vec{r})}$ 
is the same object as in Eq.~\ref{eq:Vpp} also calculated in the LSDA. By fitting the first few femtoseconds of the
demagnetization curve to a quadratic time decay we extract the demagnetization rate of the three different clusters. 
These show a systematic dependence on $\lambda_\mathrm{eff}^2$ [see Fig.~\ref{fig05}(c)]. 

In conclusion TD-SDFT calculations obtained with fully relativistic pseudopotentials for transition metal clusters 
display ultrafast demagnetization with a leading quadratic time dependence and rates ranging between 4 and 10 
$\hbar/$fs$^2$. The phenomenon is then explained in terms of the resulting laser-induced coherent drop of 
the effective magnetic field. The latter, combined with the SOC which is the strongest in the vicinity of the atomic centers, 
leads to a local decrease of the expectation 
value of $\hat{S}^z$. The external electric pulse is therefore only indirectly involved in the demagnetization process, 
that must be ascribed to the large variations of the effective magnetic field due to excited spin-polarized currents.  
Furthermore, the onset of the demagnetization shows a clear direct dependence 
on the ionic SOC properties of the material, scaling quadratically with the SOC strength. Because of the localised nature
of this ultrafast demagnetization mechanism, we believe our findings are valid beyond the cluster systems and may provide a formal 
backing to experimental observations like the comparison of Ni and CoPt$_3$ demagnetization rates in Ref. \onlinecite{Bigot09}.     

This work has been funded by the EU FP7 project CRONOS (grant no. 280879). We gratefully acknowledge the 
DJEI/DES/SFI/HEA Irish Centre for High-End Computing (ICHEC) for the provision of computational facilities 
and support and the Trinity Centre for High Performance Computing for technical support.


\bibliography{research}













\end{document}